\documentclass[conference]{IEEEtran}
\IEEEoverridecommandlockouts
\usepackage{amsmath,amsfonts}

\usepackage{algorithm}
\usepackage{array}
\usepackage[caption=false,font=normalsize,labelfont=sf,textfont=sf]{subfig}
\usepackage{textcomp}
\usepackage{stfloats}
\usepackage{url}
\usepackage{verbatim}
\usepackage{graphicx}
\usepackage{cite}
\usepackage{tabularx}
\usepackage{makecell}
\usepackage[top=2cm, bottom=2cm, left=2cm, right=2cm]{geometry}
\usepackage{algorithm}
\usepackage{algorithmicx}
\usepackage{algpseudocode}
\usepackage{amsmath}
\usepackage{amsthm}\newtheorem{myDef}{Definition}
\usepackage{makecell}
\usepackage{url}
\usepackage{cite} 
\usepackage[numbers,sort&compress]{natbib}
\usepackage{graphicx}
\usepackage{bbding}
\usepackage{graphicx}
\usepackage{subfig}
\usepackage{overpic}
\usepackage{float} 

\usepackage[misc]{ifsym}
\def\BibTeX{{\rm B\kern-.05em{\sc i\kern-.025em b}\kern-.08em
    T\kern-.1667em\lower.7ex\hbox{E}\kern-.125emX}}
\begin{document}

% \title{Topic and Stance Aware Information dissemination Model in Online Social Networks}
\title{Dynamic Information Dissemination Model Incorporating Non-Adjacent Node Interaction}
% \author{Xinyu Li, Chenwei Wang, Jinyang Huang, Xiao Sun, Meng Wang
% DM-NAI

\author{
	\IEEEauthorblockN{
		Xinyu Li\IEEEauthorrefmark{1}, 
		Jinyang Huang\IEEEauthorrefmark{1} \IEEEauthorrefmark{4},
		Xiang Zhang\IEEEauthorrefmark{2} \IEEEauthorrefmark{4},
        Peng Zhao\IEEEauthorrefmark{1},
        Meng Wang\IEEEauthorrefmark{1},
        \\Guohang Zhuang\IEEEauthorrefmark{1},
        Huan Yan\IEEEauthorrefmark{3},
		Xiao Sun\IEEEauthorrefmark{1}, 
		and Meng Wang\IEEEauthorrefmark{1}} 
	\IEEEauthorblockA{\IEEEauthorrefmark{1}School of Computer and Information, Hefei University of Technology, Hefei, China.}
	\IEEEauthorblockA{\IEEEauthorrefmark{2}CAS Key Laboratory of Electromagnetic Space Information, University of Science and Technology of China, Hefei, China.}
	\IEEEauthorblockA{\IEEEauthorrefmark{3} School of Big Data and Computer Science, Guizhou Normal University, Guiyang, China.} \IEEEauthorblockA{\IEEEauthorrefmark{4}Corresponding Authors: Jinyang Huang,  Xiang Zhang\quad Email: hjy@hfut.edu.cn, zhangxiang@ieee.org}
}

\maketitle

\begin{abstract}

Describing the dynamics of information dissemination within social networks poses a formidable challenge. Despite multiple endeavors aimed at addressing this issue, only a limited number of studies have effectively replicated and forecasted the evolving course of information dissemination. In this paper, we propose a novel model, \textit{DM-NAI}, which not only considers the information transfer between adjacent users but also takes into account the information transfer between non-adjacent users to comprehensively depict the information dissemination process. Extensive experiments are conducted on six datasets to predict the information dissemination range and the dissemination trend of the social network. The experimental results demonstrate an average prediction accuracy range of 94.62\% to 96.71\%, respectively, significantly outperforming state-of-the-art solutions. This finding illustrates that considering information transmission between non-adjacent users helps \textit{DM-NAI} achieve more accurate information dissemination predictions.

\end{abstract}


\section{Introduction}
% \IEEEPARstart{T}{he} continuous development of online social networking services has significantly facilitated people's daily lives. The ways for information sharing and communication among people are consistently expanding. However, the rapid dissemination of diverse information types inevitably results in the polarization of people's attitudes, accompanied by a significant influx of negative information, such as rumors and fake news. These phenomena give rise to a wide range of social effects that persistently manifest, including threats to the legitimacy and credibility of online platforms, disruptions to social order, impacts on the fairness of elections\cite{allcott2017social}, implications for the stock market\cite{difonzo1997rumor}, and even risks to national stability. Therefore, precisely describing the trends in information dissemination on social networking platforms has become an ongoing research focus. This is crucial for timely event comprehension and the effective management of diverse social effects.%背景，重要性
The continuous development of online social networks expands the way people share information and inevitably promotes the spread of false information, causing certain damage to social order and even affecting national stability. Therefore, precisely describing the trends in information dissemination on social network platforms has become an ongoing research focus, crucial for timely event comprehension and the effective management of diverse social effects \cite{9839133, 10278763}.

% Existing research on information dissemination models in social networks is mainly based on two classic models, the \textit{Independent Cascade (IC) Model}\cite{kempe2005influential} and the \textit{Linear Threshold (LT) Model}\cite{kempe2003maximizing}. Nonetheless, with the continuous evolution of social networking platforms, people have discovered that besides neighboring nodes, various factors such as time, topic, space, user emotions, and individual preferences also influence the process of information dissemination. Consequently, an extensive body of research has emerged, building upon the foundational \textit{IC Model}, while taking into account the influence of time\cite{guille2012predictive, haldar2023temporal, chen2012time, kim2014ct}, topic\cite{barbieri2013topic, qin2021influence, tian2020deep, michelle2016topic, zhang2020nsti}, spatial dependencies\cite{chen2017modeling}, user emotions\cite{wang2016emotion, wang2017emotion}, and individual preferences\cite{zhang2020nsti, wang2020topic, dai2022opinion}. 

Existing social network information dissemination model research is mainly based on two classic models: the \textit{Independent Cascade (IC) Model} \cite{kempe2005influential} and the \textit{Linear Threshold (LT) Model} \cite{kempe2003maximizing}. For example, based on the \textit{IC} model, researchers have carried out a large number of studies considering the influence of time \cite{haldar2023temporal,9373931, li2013modeling}, user emotions \cite{9384591,huang2021node,hung2023cecm}, topics \cite{barbieri2013topic,qin2021influence,tian2020deep,zhang2020nsti}, and individual preferences \cite{zhang2020nsti, 10339891, wang2020topic, HjyPhD, dai2022opinion}.

However, previous studies have solely focused on information transfer between adjacent users. The dissemination of information among non-adjacent users in social networks also significantly impacts the overall information dissemination. In this paper, we propose a novel information dissemination model, called \textit{DM-NAI}, which incorporates dynamic changes in user attitudes and considers the influence of non-adjacent users. By continuously updating users' attitude distribution and incorporating the information dissemination mechanism among non-adjacent users, \textit{DM-NAI} facilitates a more comprehensive prediction of the information dissemination process.

\section{Related Work}
Understanding the dissemination mechanisms behind vast amounts of information is crucial in various research fields, including viral marketing, social recommendations, community detection, and social behavior prediction. Researchers from diverse fields, such as epidemiology, computer science, and sociology, have conducted extensive studies and proposed various models of information dissemination to accurately describe and simulate this intricate process.

Kempe et al. originally introduced two fundamental models of information dissemination: the Independent Cascade Model (IC) \cite{kempe2005influential} and the Linear Threshold Model (LT) \cite{kempe2003maximizing}. Considering the influence of time, \cite{haldar2023temporal} introduced the T-IC model, effectively capturing the temporal aspects of the network. A new model called the cycle-aware intelligent method was proposed by \cite{9373931}. \cite{li2013modeling} introduced the GT dissemination model by considering the influence of time on user behavior. Accounting for the influence of user emotions, \cite{9384591} proposed the E-SFI model. \cite{huang2021node} proposed a novel model based on user sentiment. A et al. proposed CECM \cite{hung2023cecm}. Taking into account the influence of topics, \cite{barbieri2013topic} proposed a novel topic-aware influence-driven dissemination model. \cite{qin2021influence} introduced a topic-aware community-based independent cascade model. \cite{tian2020deep} presented a topic-aware social influence dissemination model. \cite{zhang2020nsti} proposed NSTI-IC model. Accounting for the influence of spatial factors, \cite{qin2021influence} introduced a community-based independent cascade model. Taking into account the influence of individual preference factors, \cite{zhang2020nsti} proposed the NSTI-IC model. \cite{wang2020topic} developed a topic-enhanced sentiment dissemination model.

% However, the above studies all ignore the role of information transfer between non-adjacent users and the dynamic change of the distribution of users' attitudes in the process of information dissemination, which greatly affects the predictive performance of the model. In contrast to these models, by fully considering the influence of the above factors, our model can make better prediction and thus improve the performance dramatically.

\section{Problem Formulation}
\begin{figure}[!t]
\centering
\includegraphics[width=2.5in]{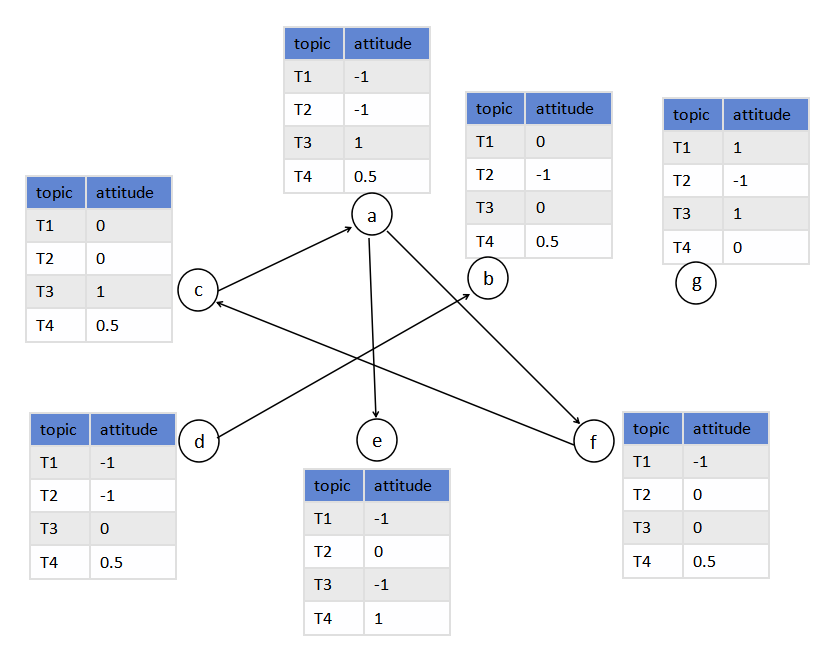}
\caption{User Network.}
\label{fig:1}
\end{figure}

% In this section, we first present the fundamental definitions used in this study and then formalize the problem that we aim to address. As illustrated in Fig.~\ref{fig:1}, the social network is modeled as a directed graph denoted as $G = (V,E,T)$, where $V$ is a node set of size $n$, $E$ is an edge set of size $m$, and $T$ is a topic set of size $z$. Where, nodes represent users in the social network, and edges represent connections between users. Each node $v_{i}\in V$ corresponds to a topic set $T_{i}=\{t_{i\_1},t_{i\_2},t_{i\_3},……,t_{i\_z}\}$, which reflects user $i$’s attitude towards information on all topics, where $t_{i\_k} \in \{-1,0,0.5, 1\}$,$t_{i\_k}$. $t_{i\_k}$ is used to represent user $i$'s attitude towards topic $k$. The value $-1$ means that user $i$ has not touched topic $k$ for the time being, that is, user $i$ is unknown to topic $k$. The value $0$ means that user $i$ has an opposing attitude towards topic $k$, the value $0.5$ means that user $i$ has a neutral attitude towards topic $k$, and the value $1$ means that user $i$ has an agreeable attitude towards topic $k$. The above three states all mean that the user $i$ is known about topic $k$.

As illustrated in Fig.~\ref{fig:1}, the social network is modeled as a directed graph denoted as $G = (V, E, T)$. Each node $v_{i}\in V$ corresponds to a topic set $T_{i}=\{t_{i\_1},t_{i\_2},t_{i\_3},……,t_{i\_z}\}$, which reflects user $i$’s attitude towards information on all topics, where $t_{i\_k} \in \{-1,0,0.5, 1\}$,$t_{i\_k}$ is used to represent user $i$'s attitude towards topic $k$. The values $-1$, $0$, $0.5$, and $1$ correspond to the user's stances, representing four states: unknown, positive, neutral, and negative. Tab.~\ref{tab1} provides an overview of frequently utilized symbols along with their corresponding interpretations.

\begin{myDef}[\textbf{Activation action}]
    Activation action can be represented by user $u$ being influenced by a sequence of information related to topic $i$ at time $t$. Consequently, there is a probability of the user's attitude towards topic $i$ changing.
\end{myDef}

% $T_i$ denotes the set of all users known to topic $i$, and at time $t$, user $u$ is influenced by a subset of users from this set. These influence mechanisms are commonly observed in various online social networks. For instance, on the Sina Weibo platform, the activation action can be described as when a user is exposed to a large amount of certain types of information, they will post a Weibo or comment on a Weibo to express their attitude.

\begin{myDef}[\textbf{Attitude distribution similarity}]
    The attitude distribution similarity between users can be described as $sim(u, v)$. Assuming that the attitude distributions of users $u$ and $v$ are $T_u=\{t_{u\_1}, t_{u\_2}, t_{u\_3}, \ldots, t_{u\_z}\}$ and $T_v=\{t_{v\_1}, t_{v\_2}, t_{v\_3}, \ldots, t_{v\_z}\}$, this value can be obtained by analyzing the similarity of the distributions of $T_u$ and $T_v$.
\end{myDef}

\begin{myDef}[\textbf{Information dissemination}]
    We define the conditions for information dissemination in social networks as follows:$(1) (u,v) \in E$; $(2) (u,v) \notin E$, but $sim(u,v) \textgreater \tau$. Here,$\tau$ represents the threshold for controlling information dissemination. When either condition $(1)$ or condition $(2)$ is satisfied, there is a certain probability of information dissemination occurring between users $u$ and $v$.
\end{myDef}

% Obviously, when a piece of information spreads from user $u$ to $v$, there must be a social connection or interest association between these users. We use attitude distribution similarity to quantify the strength of interest association between users, and it is defined as follows:

% The greater the attitude distribution similarity between users $u$ and $v$, the more likely it is for information to be transferred between them, and the greater the potential for mutual influence between the two users.

\begin{myDef}[\textbf{Social influence}]
    For user $u$ and user $v$ in a social network, we define $P_{i}(u, v)$ to represent the influence of user $u$ on user $v$'s attitude regarding topic $i$.
\end{myDef}

% \begin{myDef}[\textbf{dissemination cascade}]
%     For any user $u$, $v$ of the social network, when user $v$ is activated by user $u$ at a certain moment, user $v$ is contagious and cannot be deactivated. In the subsequent process of information dissemination, user $v$ can influence other users according to the Definition $2$.
% \end{myDef}

\begin{myDef}[\textbf{User attitude's perseverance}]
    For user $v$, we define $A_{v}^{i}$ to represent the perseverance of user $v$'s attitude towards topic $i$, which is the probability of change in user $v$'s attitude towards topic $i$.
\end{myDef}

\section{The Proposed Model}

\begin{table*}[t]
\caption{Important Mathematical Symbols.\label{tab1}}
\centering
\begin{tabularx}{\textwidth}{ c
         >{\raggedright\arraybackslash}X
        }
\hline
Notation & Description \\
\hline
$G=(\boldsymbol{V},\boldsymbol{E},\boldsymbol{T})$ & Represents a social network, where $\boldsymbol{V}$ is a set of nodes with a size of n, $\boldsymbol{E}$  is a set of edges with a size of m \cite{chen2021information}, and $\boldsymbol{T}$  is a set of topics with a size of z.\\

$P_e$ & Probability of information dissemination on edge e \cite{wu2020collaborative}.\\

$n = |\boldsymbol{V}|$ & The number of nodes in G.\\

$m =|\boldsymbol{E}|$ & The number of edges in G\cite{10251628}.\\

$\boldsymbol{T_i}=<t_{i\_1},t_{i\_2},……,t_{i\_z}>$ & Attitude held by user i towards information on all topics.\\

$z = |\boldsymbol{T}|$ & The number of topics in $G$.\\

$\delta, \lambda, \mu, \epsilon$ & Control parameter.\\

\hline
\end{tabularx}
\end{table*}

% We propose a novel information dissemination model in social networks. By introducing the evaluation mechanism of users' attitude changes in the process of information dissemination, the model can capture the dynamic characteristics of users' attitudes in the information dissemination process. In addition, based on the user's attitude distribution, we introduce an evaluation mechanism for information dissemination between non-adjacent users, so as to comprehensively describe the information dissemination process of the network. In our model, the information dissemination process unfolds in discrete time steps $t$, and starts from an initial set of active users. When user $v$ receives information about topic $i$ at a certain moment, by comprehensively considering all the information that the user has received so far about the topic $i$, it can be determined whether the user's attitude towards the topic will change at the next moment. If the attitude of user $v$ changes, then in the next process of information dissemination, the user will influence other users with a new attitude. We now describe the proposed model in more detail. For better illustration, Tabel~\ref{tab1} lists some mathematical symbols used in this paper.

\subsection{Attitude Distribution Similarity}
% In the social network model of this paper, user is depicted as the distribution of the user's attitudes towards various topics. The similarity of attitude distribution among users affects the probability of information dissemination among users. Given two nodes $u$, $v$, the topic sets corresponding to these two nodes are $\boldsymbol{T_{u}}=\{t_{u\_1},t_{u\_2},t_{u\_3},……, t_{u\_z}\}$ and $\boldsymbol{T_{v}}=\{t_{v\_1},t_{v\_2},t_{v\_3},……,t_{v\_z}\}$. The topic similarity $sim(u,v)$ between node $u$ and node $v$ is measured as follows:
In the social network model of this paper, the user is depicted as the distribution of attitudes towards various topics. The similarity \cite{huang2020towards} of attitude distribution among users affects the probability of information dissemination. Given two nodes $u$ and $v$, the topic similarity $sim(u,v)$ between nodes $u$ and $v$ is measured as follows:

\begin{equation}
sim(u,v) = 1/\bigg( \Big( \sqrt{z}+\sqrt{\sum_{i=1}^z\big( t_{u}^{i} - t_{v}^{i}\big)^2}\Big)/\sqrt{z}\bigg) \label{eq:1}
\end{equation}
where a higher value of $sim(u, v)$ indicates greater similarity in the attitude distribution between nodes $u$ and $v$, leading to a higher probability of information dissemination between them.

\subsection{Influence Evaluation between Users}
In a social network, there exists a corresponding influence between users when they transmit information \cite{zhang2023wital}. Users can leverage this influence to assess whether and how their attitudes will change. We define $r_{uv}^{i}$ as the dissemination rate between users $u$ and $v$ regarding topic $i$. A larger value denotes a higher rate of information dissemination.

Assuming that user $u$ transmits information about topic $i$ to user $v$ within time $\tau_i$, the probability that the information cannot be transferred from $u$ to $v$ can be obtained by multiplying the probability of non-transfer in each small time interval $\triangle t$, that is

\begin{equation}
\begin{aligned}
P_{i}(u,v) =(1 - W_{i}(u,v)) \ast sim(u,v) \ast f(t_{v}^{i},t_{u}^{i})\label{eq:2}
\end{aligned}
\end{equation}

\begin{equation}
\begin{aligned}
W_{i}(u,v) =1- \lim_{\triangle t \to 0} \big(1 - r_{uv}^{i} \triangle t \big)^{\tau_i/\triangle t} = 1 - e^{-r_{uv}^{i}t}\label{eq:3}
\end{aligned}
\end{equation}

\begin{equation}
f(t_{v}^{i},t_{u}^{i})=\left\{
	\begin{array}{rcl}
	1 ,& & t_{v}^{i} = -1,0.5 or t_{v}^{i} = t_{u}^{i}\\
	\lambda, & & |t_{v}^{i} - t_{u}^{i}| \le 0.5\\
	\mu ,& & else\\
	\end{array}
	\right.   
\label{eq:4}
\end{equation}
% where $\lambda$ and $\mu$ are control parameters used to regulate the influence weight between user $u$ and user $v$ regarding topic $i$. Irrespective of whether there exists a connection relationship between two users and how similar the attitude distribution between the two users is, there is a certain probability of mutual influence. This differs from previous models that solely considered information dissemination between adjacent users. In cases where the similarity of attitude distributions between non-adjacent users exceeds that between adjacent users, the influence between non-adjacent users may even surpass that between adjacent users.
where $\lambda$ and $\mu$ are control parameters used to regulate the influence weight between user $u$ and user $v$ regarding topic $i$. Regardless of the existence of a connection relationship between two users and the similarity of the attitude distribution between the two users, there is a certain probability of mutual influence.

\subsection{Attitude Change Probability}
% Considering the different attitudes of users on topics and the intricacies of the information received, the probability of changes in users' attitudes may be different. We define $A_{v}^{i}$ as the attitude perseverance of user $v$ towards topic $i$, which serves as a metric to gauge the likelihood of the user's attitude change. A higher value of $A_{v}^{i}$ indicates that user $v$ is less likely to change his attitude at that moment.
Considering the different attitudes of users on topics and the intricacies of the information received, the probability of changes in users' attitudes may vary. We define $A_{v}^{i}$ as the attitude perseverance of user $v$ towards topic $i$, which serves as a metric to gauge the likelihood of the user's attitude change. $A_{v}^{i}$ is calculated as follows:

% Assuming that at time $t$, user $v$ has received a total of $k$ pieces of information about topic $i$. The attitude perseverance of user $v$ towards topic $i$ at this moment can be defined as follows:

\begin{equation}
A_{v}^{i} = A_{v}^{i} - \sum_{u=1}^k\Big(|t_{u}^{i} - t_{v}^{i}| \ast P_{i}(u,v) - (\overline{t_{u}^{i} \oplus t_{v}^{i}}) \ast P_{i}(u,v) \Big)/k
\label{eq:5}
\end{equation}
% the attitude perseverance $A_{v}^{i}$ of user $v$ towards topic $i$ is influenced by various types of information he receives. When user $v$ encounters information that conflicts with his own attitude, $A_{v}^{i}$ is reduced, indicating an increased possibility of attitude change towards topic $i$. Conversely, when user $v$ receives information that aligns with his existing attitude on topic $i$, $A_{v}^{i}$ increases, indicating a decreased possibility of attitude change for user $v$ towards topic $i$.
when user $v$ encounters information that conflicts with his attitude, $A_{v}^{i}$ is reduced, indicating an increased possibility of attitude change towards topic $i$. Conversely, when user $v$ receives information that aligns with his existing attitude on topic $i$, $A_{v}^{i}$ increases, indicating a decreased possibility of attitude change for user $v$ towards topic $i$.

\subsection{Attitude Change Mechanism}
% In social networks, changes in a user's attitude towards a particular topic are determined by the user's own attitude and the information that the user has received. Given a user $v$, suppose user $v$ is unknown to the topic $i$ at the moment $t$. At this point, user $v$ receives information from user $u$, and user $u$ has his own attitude towards topic $i$. The mechanism for user $v$'s attitude change at the next moment is as follows:
In social networks, changes in a user's attitude towards a particular topic are determined by the user's attitude and the information that the user has received. The mechanism for user $v$'s attitude change at the next moment is as follows:

\begin{equation}
t_{v}^{i}=\left\{
	\begin{array}{rcl}
	t_{u}^{i} ,& & P_{i}(u,v)\ge A_{v}^{i}\\
	0.5, & & else\\
	\end{array}
	\right.   
\label{eq:6}
\end{equation}
that is if user $v$ has not been exposed to relevant information about the topic $i$ before time $t$, the change in user $v$'s attitude at the next moment is determined by the similarity of attitude distributions between users and the perseverance of user $v$'s attitude towards topic $i$. The higher the similarity between the attitude distributions, the greater the likelihood that user $v$ will be influenced by the other user's attitude.

When user $v$ has been exposed to relevant information about the topic before time $t$, the mechanism for the user's attitude change at the next moment is as follows:

\begin{equation}
t_{v}^{i} = (\overline{t_{u}^{i} \oplus t_{v}^{i}}) \ast  t_{v}^{i} + (t_{u}^{i} \oplus t_{v}^{i}) \ast (t_{v}^{i} \pm \varepsilon \ast 0.5)
\label{eq:7}
\end{equation}
% where $\varepsilon$ is a constant parameter, when $P_{i}(u,v) \textgreater A_{v}^{i}$, $\varepsilon$ is 1, when $P_{i}(u,v) \textless A_{v}^{i}$,$\varepsilon$ is 0,$\pm$ is used to control the direction of stance change of node v. When the initial $t_{v}^{i}$ is 0, indicating that node v initially holds an opposing stance on topic $t_i$, we assign a positive sign (+). Conversely, when the initial $t_i$ is 1, indicating that node v initially holds a supportive stance on topic $t_i$, we assign a negative sign (-).
where $\varepsilon$ is a constant parameter. When $P_{i}(u,v) \textgreater A_{v}^{i}$, $\varepsilon$ is $1$. When $P_{i}(u,v) \textless A_{v}^{i}$,$\varepsilon$ is $0$. The symbol $\pm$ is used to control the direction of the stance change of node $v$. When the initial $t_{v}^{i}$ is $0$, indicating that node $v$ initially holds an opposing stance on topic $t_i$, we assign a positive sign (+). Conversely, we assign a negative sign (-).

\floatname{algorithm}{Algorithm}
\renewcommand{\algorithmicrequire}{\textbf{Input:}}
\renewcommand{\algorithmicensure}{\textbf{Output:}}

\begin{algorithm}
    \caption{DM-NAI}
        \begin{algorithmic}[1] %每行显示行号
            \Require $G=(\boldsymbol{V},\boldsymbol{E}, \boldsymbol{T})$, initial set of topics $\boldsymbol{T_{k}}=\{\boldsymbol{T_{1}},\boldsymbol{T_{2}},……,\boldsymbol{T_{z}}\}$,initial topic number j,number of dissemination K.
            \Ensure  The updated network G=(V,E,T).
            \Function {DM-NAI}{$G,\boldsymbol{T_{k}}, j, K$}
            \State $\boldsymbol{T_{j-0}^{now}} = \boldsymbol{T_{j-0}}$, $\boldsymbol{T_{j-0.5}^{now}} = \boldsymbol{T_{j-0.5}}$, $\boldsymbol{T_{j-1}^{now}} = \boldsymbol{T_{j-1}}$
            \State $\boldsymbol{V_{j}^{new}} = \boldsymbol{T_{j-0}^{now}} \cup \boldsymbol{T_{j-0.5}^{now}} \cup \boldsymbol{T_{j-1}^{now}}$
            \State $\boldsymbol{V_{adj}} = \boldsymbol{\emptyset}$, $\boldsymbol{S} = \boldsymbol{\emptyset}$
                \For{k = 1 to K}
                    \For{v in $\boldsymbol{V_{j}^{new}}$}
                        \State $\boldsymbol{S}=\{u|(u,v)\in \boldsymbol{E}\}$
                        \For{q $\in$ $\boldsymbol{S}$}
                            \If{q $\notin \boldsymbol{V_{adj}}$}
                                \State $T_{cur} = T_{q}^{j}$
                                \State $A_{q}^{j} = Eq.(4)$
                                \State $T_{q}^{j} = ATT(G,q,v,j,A_{q}^{j})$
                                \State $\boldsymbol{V_{adj}} \gets q$
                            \EndIf
                            \If{$T_{cur} = -1$ and $T_{q}^{j} != -1$}
                                \State $\boldsymbol{T_{j-T_{q}^{j}}} \gets q$
                                \State $\boldsymbol{V_{j}^{new}} \gets q$
                            \EndIf
                        \EndFor
                    \EndFor
                    \State $\boldsymbol{V_{\sim adj}} \gets \boldsymbol{V}\backslash \boldsymbol{V_{adj}}$
                    \State $NADJ(G,j,\boldsymbol{V_{\sim adj}},\boldsymbol{V_{j}^{new}})$
                \EndFor
                \State \Return{$G$}
            \EndFunction
        \end{algorithmic}
\end{algorithm}

\floatname{algorithm}{Algorithm}
\renewcommand{\algorithmicrequire}{\textbf{Input:}}
\renewcommand{\algorithmicensure}{\textbf{Output:}}

\begin{algorithm}
    \caption{ATT}
        \begin{algorithmic}[1] %每行显示行号
            \Require $G=(\boldsymbol{V},\boldsymbol{E}, \boldsymbol{T})$, node q,v, the topic number j,$A_{q}^{j}$.
            \Ensure  $T_{q}^{j}$.
            \Function {ATT}{$G,q,v,j,A_{q}^{j}$}
            \State $T_{cur} = T_{q}^{j}$
             \If{$T_{q}^{j}=0.5$ or -1}
                \State $T_{q}^{j} = Eq.(5)$
            \Else
                \State $T_{q}^{j} = Eq.(6)$
            \EndIf
            \If{$T_{cur} != T_{q}^{j}$}
                \State $\boldsymbol{T_{j-T_{q}^{j}}} \gets q$
                \State $\boldsymbol{T_{j-T_{cur}}} \gets q$
            \EndIf
            \State \Return{$T_{q}^{j}$}
            \EndFunction
        \end{algorithmic}
\end{algorithm}

\section{Algorithm}
% Unlike previous algorithms, the model proposed in this paper is insensitive to time and considers the influence of non-neighboring nodes. Additionally, it incorporates the calculation of user stances on topics, taking into account the impact of user stances on the dissemination process. The detailed procedure of the algorithm is as follows.

\subsection{Information dissemination Process}
% Different from traditional information dissemination models, we propose a Topic and Stance Aware social network dissemination model based on the current landscape of new-generation social networks. Considering that information in the current landscape of new-generation social networks is no longer time-sensitive, meaning that users are exposed to not only the most recent information but also information from previous time points, our model assumes that the information posted by user v at any time $t_{1}$ may be received by user u at a later time $t_{2}$ (where $t_{2}(t_{2} \textgreater t_{1})$). In addition, for users in a social network, the information they are exposed to is not limited to their connected nodes. Therefore, our model assumes that user v can access information not only from its neighboring nodes but also from non-neighboring nodes. Regardless of the topic similarity sim(u,v) between users, user v has a certain probability of accessing information from other users u. The specific dissemination process is illustrated in Algorithm $1$.

Our model assumes that user $v$ can access information not only from its neighboring nodes but also from non-neighboring nodes. Additionally, the model assumes that the information posted by user $v$ at any time $t_{1}$ may be received by user $u$ at a later time $t_{2}$ (where $t_{2} \textgreater t_{1}$). The specific dissemination process is illustrated in Algorithm $1$.

% $\boldsymbol{V}\backslash \boldsymbol{V_{adj}}$ denotes the difference set between the set $\boldsymbol{V}$ and the set $\boldsymbol{V_{adj}}$, for example, $\boldsymbol{V}=\{a,b,c,d,e\}$, $\boldsymbol{V_{adj}}=\{b,d\}$, then $\boldsymbol{V}\backslash \boldsymbol{V_{adj}} =\{a, c,e\}$. In the DM-NAI algorithm, the adjacency node $\boldsymbol{V_{adj}}$ and the non-adjacency node $\boldsymbol{V_{~adj}}$ are comprehensively considered, as well as all the users known and unknown to the specified topic j.
$\boldsymbol{V}\backslash \boldsymbol{V_{adj}}$ denotes the difference set between the set $\boldsymbol{V}$ and the set $\boldsymbol{V_{adj}}$. In the algorithm $1$, the adjacency node $\boldsymbol{V_{adj}}$ and the non-adjacency node $\boldsymbol{V_{\sim adj}}$ are comprehensively considered.

\subsection{Node State Change Process}
% During the process of information dissemination, the stance of a node towards the current topic may continuously change. We consider the current stance of the node and the influence of all the different stance information it has received up until time t to determine the possible stance changes at the current time t. Clearly, when the node is in an unknown or neutral state, it is susceptible to the influence of different stance information. However, when the node is in a supporting or opposing state for the current topic, the change in stance depends on factors such as the similarity of topics between the node and other users, as well as all the different stance information it has received so far. The specific calculation process is shown in Algorithm $2$.
During the information dissemination process, the stance of a node towards the current topic may continuously change. We consider the node's current stance and the influence of all the different stance information it has received up until time $t$ to determine the possible stance changes at the current time $t$. The specific calculation process is shown in Algorithm $2$.

% The process of state change for node $q$ regarding topic $j$ can be determined based on Definition $4$. It distinguishes the state of node $q$ for a specific topic $j$. If node $q$ is in an unknown or neutral state for topic $j$, it can be derived using Eq.~\ref{eq:5}. If node $q$ is in a supporting or opposing state for topic $j$, it can be derived using Eq.~\ref{eq:6}. Particularly, when the state of node $q$ for topic $j$ changes, the corresponding set also undergoes dynamic changes.

\subsection{dissemination in Non-adjacent Nodes}
Nodes also influence their non-neighboring nodes during the information dissemination process. Whether these nodes are in a known or unknown state for the current topic, there is a certain probability that they will be influenced by the information from the node. The specific dissemination process is shown in Algorithm $3$.

\floatname{algorithm}{Algorithm}
\renewcommand{\algorithmicrequire}{\textbf{Input:}}
\renewcommand{\algorithmicensure}{\textbf{Output:}}

\begin{algorithm}
    \caption{NADJ}
        \begin{algorithmic}[1] %每行显示行号
            \Require $G=(\boldsymbol{V},\boldsymbol{E}, \boldsymbol{T})$,$\boldsymbol{V_{~adj}},\boldsymbol{V_{j}^{new}}$,the number of dissemination k, the topic number j, the parameters r1,r2.
            \Ensure  The updated network $G=(\boldsymbol{V},\boldsymbol{E},\boldsymbol{T})$.
            \Function {NADJ}{$G,\boldsymbol{V_{~adj}},\boldsymbol{V_{j}^{new}},k,j,r1,r2$}
            \State Generating collections $\boldsymbol{V^{new}},\boldsymbol{V_{~adj}^{new}},\boldsymbol{V^{new}}=\{v|v\in \boldsymbol{V_{j}^{new}}\}$
            \State $|\boldsymbol{V^{new}}|=r1\ast|\boldsymbol{V_{j}^{new}}|$
            \State $\boldsymbol{V_{~adj}^{new}}=\{\boldsymbol{U} \cup \boldsymbol{V}|u,v \in \boldsymbol{V_{~adj}},u \in \boldsymbol{T_{j}}, v \notin \boldsymbol{T_{j}}\}$
            \State $|\boldsymbol{V_{~adj}^{new}}|=r2\ast|\boldsymbol{V_{~adj}}|$
            \For{$q \in \boldsymbol{V_{~adj}^{new}}$}
                \State $T_{cur} = T_{q}^{j}$
                \For{$v \in \boldsymbol{V^{new}}$}
                    \State $A_{q}^{j} = Eq.(4)$
                    \State $ATT(G,q,v,j,A_{q}^{j})$
                    \If{$T_{cur} = -1$ and $T_{q}^{j} != -1$}
                        \State $\boldsymbol{V_{j}^{new}} \gets q$
                        \State $\boldsymbol{T_{j}} \gets q$
                    \EndIf
                \EndFor
            \EndFor
            \State \Return{$G$}
            \EndFunction
        \end{algorithmic}
\end{algorithm}
% where, $\boldsymbol{T_j}$ represents the set of users for whom topic j is known, and $\boldsymbol{V_{~adj}^{new}}$  contains a subset of non-neighboring users for node v. The parameter r2 determines the size of the set $\boldsymbol{V_{~adj}^{new}}$. And for the collection of certain proportion relationship in the $\boldsymbol{U}$ and $\boldsymbol{V}$, the $|\boldsymbol{U}| = r \ast |\boldsymbol{V_{~adj}^{new}}|$, $|\boldsymbol{V}| = a \ast |\boldsymbol{V_{~adj}^{new}}|$, among them, $r\in[0.5,1], a\in[0,0.5),r + a = 1$.
where, $r\in[0.5,1], a\in[0,0.5), r + a = 1$. $\boldsymbol{T_j}$ represents the set of users for whom topic $j$ is known, and $\boldsymbol{V_{~adj}^{new}}$  contains a subset of non-neighboring users for node $v$.

\section{Experiment}
% In this section, we will perform tests on our algorithm using six real-world datasets obtained through Python web scraping. To showcase the effectiveness and credibility of our dissemination method, we will compare it with several existing models. Additionally, we will conduct visual analysis of user activation and stance changes during the information dissemination process.

\subsection{Dataset}

The datasets utilized in the experiments of this paper are sourced from actual microblog data. Each dataset features varying numbers of topics and microblog posts. Tab.~\ref{tab2} provides detailed information about the datasets.

\begin{table}[!t]
\caption{The Dataset.\label{tab2}}
\begin{center}
\begin{tabular}{c c c c c}
\hline
\textbf{Dataset} & \textbf{nodes} & \textbf{edges} & \textbf{topics} & \textbf{Number of seed nodes} \\
\hline

\textbf{Dataset \uppercase\expandafter{\romannumeral1}}  & 1331 & 8737 & 1 & 20\\

\textbf{Dataset \uppercase\expandafter{\romannumeral2}} & 1109 & 7723 & 1 & 23\\

\textbf{Dataset \uppercase\expandafter{\romannumeral3}} & 1801 & 8493 & 1 & 27\\

\textbf{Dataset \uppercase\expandafter{\romannumeral4}} & 2351 & 14739 & 2 & 41\\

\textbf{Dataset \uppercase\expandafter{\romannumeral5}} & 2817 & 13283 & 2 & 45\\

\textbf{Dataset \uppercase\expandafter{\romannumeral6}} & 4028 & 23151 & 3 & 69\\
\hline
\end{tabular}
\end{center}
\end{table}

\begin{figure*}[t]
	\centering
	\subfloat[dissemination process of dataset \uppercase\expandafter{\romannumeral1}.]{\includegraphics[width=0.33\textwidth]{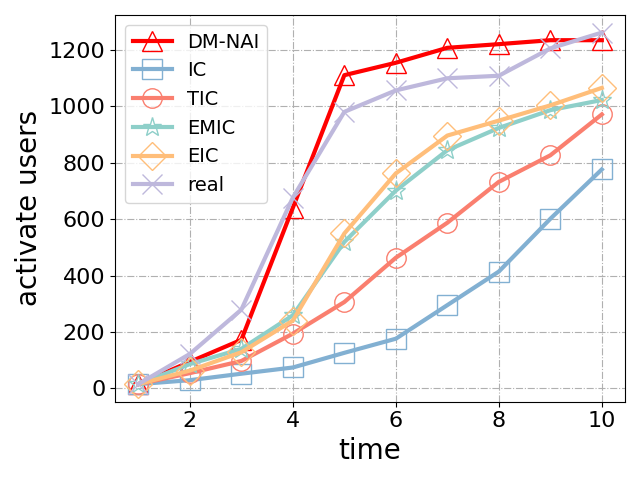}}
	\subfloat[dissemination process of dataset \uppercase\expandafter{\romannumeral2}.]{\includegraphics[width=0.33\textwidth]{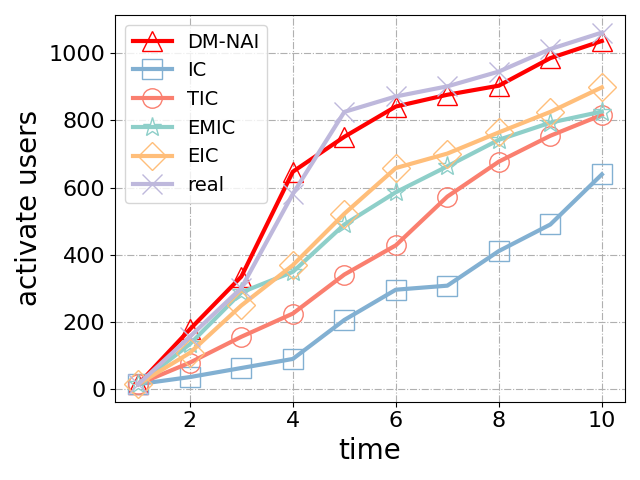}}
	\subfloat[dissemination process of dataset \uppercase\expandafter{\romannumeral3}.]{\includegraphics[width=0.33\textwidth]{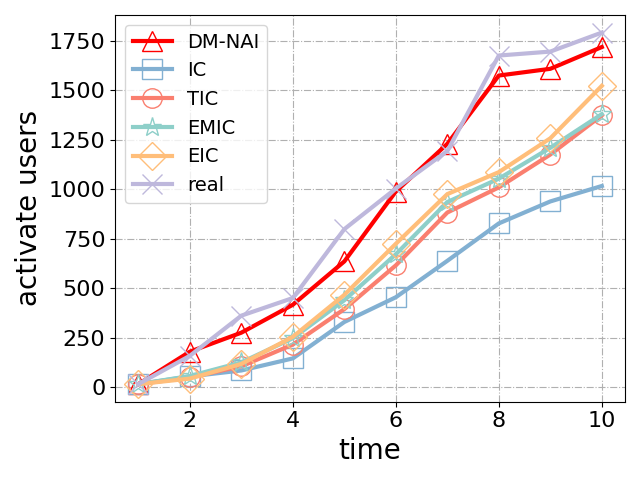}}\\
    \subfloat[dissemination process of dataset \uppercase\expandafter{\romannumeral4}.]{\includegraphics[width=0.33\textwidth]{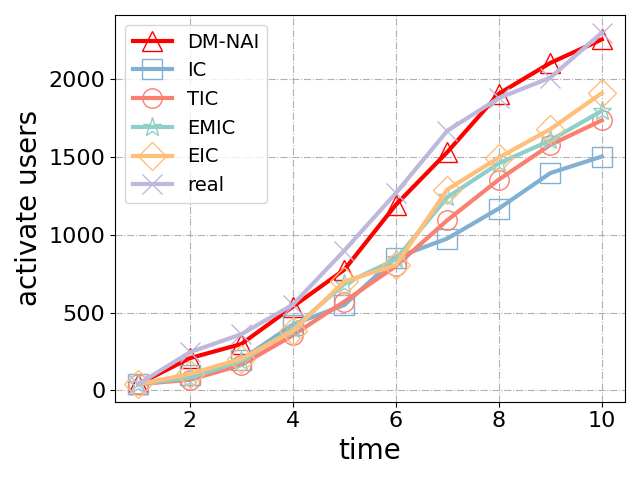}}
	\subfloat[dissemination process of dataset \uppercase\expandafter{\romannumeral5}.]{\includegraphics[width=0.33\textwidth]{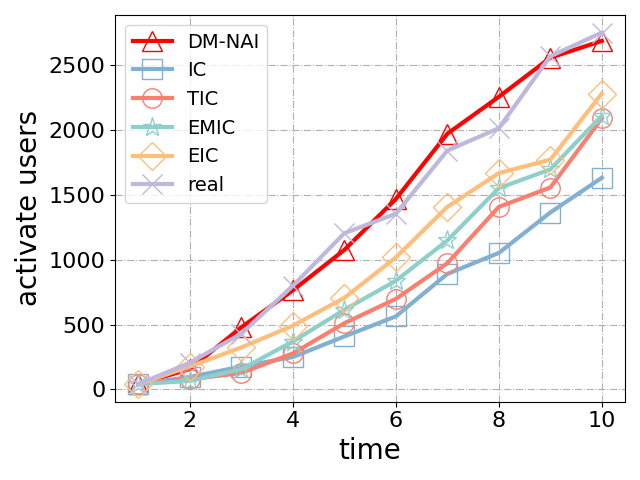}}
	\subfloat[dissemination process of dataset \uppercase\expandafter{\romannumeral6}.]{\includegraphics[width=0.33\textwidth]{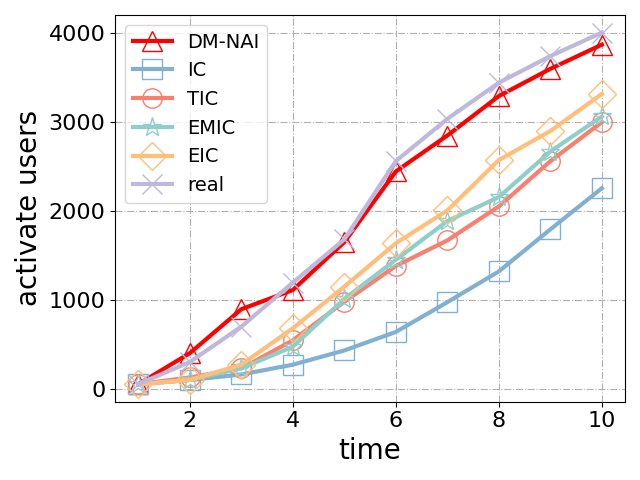}}
	\caption{The dissemination process from dataset \uppercase\expandafter{\romannumeral1} to dataset \uppercase\expandafter{\romannumeral6}.}
    \label{fig2}
\end{figure*}

% \begin{figure*}[t]
%   \centering
%   \subfloat[Stance accuracy for dataset \uppercase\expandafter{\romannumeral1}.]
%   {\includegraphics[width=0.3\textwidth]{F1-att.png}\label{f:fig8}}
%   [b]    
%   \subfloat[Stance accuracy for dataset \uppercase\expandafter{\romannumeral2}.]
%   {\includegraphics[width=0.3\textwidth]{F2-att.png}\label{f:fig9}}
%   [b]
%   \subfloat[Stance accuracy for dataset \uppercase\expandafter{\romannumeral3}.]
%   {\includegraphics[width=0.3\textwidth]{F3-att.png}\label{f:fig10}}
%   [b]
%   \caption{Stance accuracy from dataset \uppercase\expandafter{\romannumeral1} to dataset \uppercase\expandafter{\romannumeral3}}
%   \label{fig3}
% \end{figure*}

\subsection{Analysis of dissemination Results}

Based on the six datasets, we conduct experiments to analyze the accuracy of the model. A comparison is made with the IC model, as well as existing models based on user sentiment or topic, namely TIC \cite{barbieri2013topic}, EMIC \cite{wang2016emotion}, and EIC \cite{dai2022opinion}. The experimental results are presented in Tab.~\ref{tab3}. The results demonstrate that the proposed model outperforms existing models in terms of accuracy. Furthermore, Fig.~\ref{fig2} illustrates the dynamic changes of the affected users when information is propagated through the DM-NAI, IC, TIC, EMIC, and EIC models, respectively, for each of the six datasets. The similarity between our model and the real information dissemination process is further highlighted by comparing it with the real dissemination process in social networks.

\begin{table}[!t]
\caption{AUC values for IC, LT, EMIC, EIC, TIC and DM-NAI.\label{tab3}}
\centering
\begin{tabular}{c c c c c c}
\hline
Dataset & IC & TIC & EMIC & EIC & DM-NAI \\
\hline
Dataset \uppercase\expandafter{\romannumeral1} & 61.54$\%$ & 77.01$\%$ & 78.63$\%$ & 84.48$\%$ & 95.90$\%$ \\

Dataset \uppercase\expandafter{\romannumeral2} & 61.50$\%$ & 76.92$\%$ & 77.92$\%$ & 84.64$\%$ & 96.71$\%$ \\

Dataset \uppercase\expandafter{\romannumeral3} & 59.10$\%$ & 76.68$\%$ & 77.01$\%$ & 84.97$\%$ & 95.82$\%$ \\

Dataset \uppercase\expandafter{\romannumeral4} & 65.35$\%$ & 75.49$\%$ & 78.15$\%$ & 83.12$\%$ & 95.01$\%$\\

Dataset \uppercase\expandafter{\romannumeral5} & 59.30$\%$ & 75.97$\%$ & 76.69$\%$ & 82.89$\%$ & 94.67$\%$\\

Dataset \uppercase\expandafter{\romannumeral6} & 56.28$\%$ & 74.82$\%$ & 76.43$\%$ & 82.77$\%$ & 94.62$\%$ \\
\hline
\end{tabular}
\end{table}

% \begin{figure}[!t]
% \centering
% \includegraphics[width=2.5in]{F1-diff.png}
% \caption{dissemination process of dataset \uppercase\expandafter{\romannumeral1}.}
% \label{fig2}
% \end{figure}

% \begin{figure}[!t]
% \centering
% \includegraphics[width=2.5in]{F2-diff.png}
% \caption{dissemination process of dataset \uppercase\expandafter{\romannumeral2}.}
% \label{fig3}
% \end{figure}

% \begin{figure}[!t]
% \centering
% \includegraphics[width=2.5in]{F3-diff.png}
% \caption{dissemination process of dataset \uppercase\expandafter{\romannumeral3}.}
% \label{fig4}
% \end{figure}

% \begin{figure}[!t]
% \centering
% \includegraphics[width=2.5in]{F12-diff.png}
% \caption{dissemination process of dataset \uppercase\expandafter{\romannumeral4}.}
% \label{fig5}
% \end{figure}

% \begin{figure}[!t]
% \centering
% \includegraphics[width=2.5in]{F23-diff.png}
% \caption{dissemination process of dataset \uppercase\expandafter{\romannumeral5}.}
% \label{fig6}
% \end{figure}

% \begin{figure}[!t]
% \centering
% \includegraphics[width=2.5in]{F123-diff.png}
% \caption{dissemination process of dataset \uppercase\expandafter{\romannumeral6}.}
% \label{fig7}
% \end{figure}

\begin{figure*}[t]
	\centering
	\subfloat[Stance change for dataset \uppercase\expandafter{\romannumeral1}.]{\includegraphics[width=0.33\textwidth]{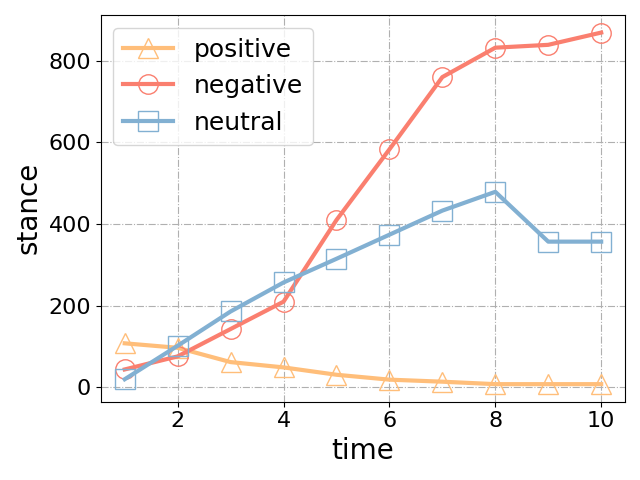}}
	\subfloat[Stance change for dataset  \uppercase\expandafter{\romannumeral2}.]{\includegraphics[width=0.33\textwidth]{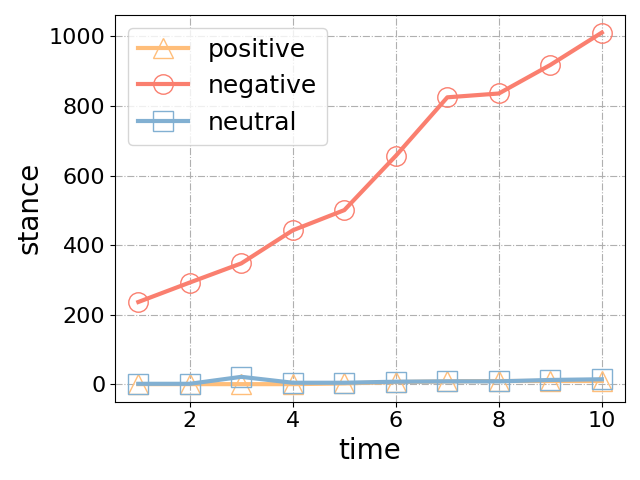}}
	\subfloat[Stance change for dataset \uppercase\expandafter{\romannumeral3}.]{\includegraphics[width=0.33\textwidth]{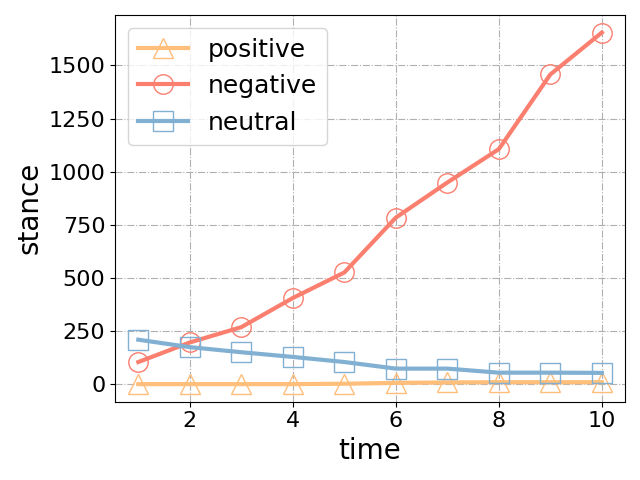}}
	\caption{The stance change process from dataset \uppercase\expandafter{\romannumeral1} to dataset \uppercase\expandafter{\romannumeral3}.}
    \label{fig3}
\end{figure*}

\begin{table}[!t]
\caption{Stance prediction accuracy.\label{tab4}}
\centering
\begin{tabular}{c c}
\hline
Dataset    & Stance prediction accuracy \\
\hline
Dataset \uppercase\expandafter{\romannumeral1}  & 79.40$\%$\\

Dataset \uppercase\expandafter{\romannumeral2} & 76.92$\%$ \\

Dataset \uppercase\expandafter{\romannumeral3} & 77.41$\%$\\
\hline
\end{tabular}
\end{table}

\subsection{Stance Change Analysis}
% In this paper, Experimental data sets\uppercase\expandafter{\romannumeral1}, \uppercase\expandafter{\romannumeral2}, and \uppercase\expandafter{\romannumeral3}, are used to analyze change of users' stance, and the experimental results are shown in Tabel~\ref{tab4}. The experimental results show that our model also has high accuracy in predicting the user's stance status on the topic in information dissemination. In addition, we also conducted a dynamic analysis of the stance changes in the above three datasets during the dissemination of information dissemination, as shown in Fig.~\ref{fig3}. Obviously, in the process of information dissemination, users with different positions will continue to change, and due to the promotion of the network, users tend to have a negative attitude toward the topic of public opinion. With the continuous dissemination of information, users who were originally in a supportive or neutral position may also be influenced to change to a negative attitude.

In this paper, experimental data sets\uppercase\expandafter{\romannumeral1}, \uppercase\expandafter{\romannumeral2}, and \uppercase\expandafter{\romannumeral3}, are used to analyze change of users' stance, and the experimental results are shown in Tab.~\ref{tab4}.

% The results indicate that our model achieves high accuracy in predicting users' stances towards the topics during the information dissemination process. Furthermore, dynamic analysis of stance changes in the information spreading process was conducted for the aforementioned three datasets, as depicted in Fig.~\ref{fig3}. It is evident that users with different stances undergo continuous changes during the information dissemination process. Additionally, due to the influence of network dynamics, users tend to adopt a negative attitude towards public opinion topics. Moreover, as information spreads, users who initially held a supportive or neutral stance may also be influenced and shift towards a negative stance.

The results indicate that users with different stances undergo continuous changes during the information dissemination process. Additionally, influenced by network dynamics, users tend to adopt a negative attitude towards public opinion topics. Moreover, as information spreads, users who initially held a supportive or neutral stance may also be influenced and shift towards a negative stance.

% \begin{figure}[!t]
% \centering
% \includegraphics[width=2.5in]{F1-att.png}
% \caption{Stance accuracy for dataset \uppercase\expandafter{\romannumeral1}.}
% \label{fig8}
% \end{figure}

% \begin{figure}[!t]
% \centering
% \includegraphics[width=2.5in]{F2-att.png}
% \caption{Stance accuracy for dataset \uppercase\expandafter{\romannumeral22}.}
% \label{fig9}
% \end{figure}

% \begin{figure}[!t]
% \centering
% \includegraphics[width=2.5in]{F3-att.png}
% \caption{Stance accuracy for dataset \uppercase\expandafter{\romannumeral3}.}
% \label{fig10}
% \end{figure}

\section{CONCLUSION}
% In the context of information dissemination in online social networks, traditional information dissemination models have become inadequate to adapt to the current network landscape, which involves a lot of information transmission among non-adjacent users. Therefore, this paper proposes a novel Dynamic Model (DM-NAI)for social network information dissemination and dissemination. To validate the proposed model, experiments were conducted on six real-world datasets. The experimental results demonstrate that the proposed algorithm in this paper outperforms traditional information dissemination models in simulating the information dissemination process in current social networks. Furthermore, this paper analyzes the changes in users' stances towards topics during the information dissemination process and dynamically illustrates the process of stance changes among users throughout the information spread. In future research, we will explore methods to select initial seed nodes that achieve the maximum dissemination range while minimizing the budget requirements. Additionally, we will seek methods to promptly intercept the spread of negative information, aiming to minimize the occurrence of adverse events in the network.
In this paper, we propose a novel model, \textit{DM-NAI}, which considers information transfer between non-adjacent users. By continuously calculating the similarity of changing attitude distribution and the probability that one node influences another non-adjacent node, \textit{DM-NAI} effectively incorporates the impact of user attitude changes and interactions with non-adjacent nodes on information dissemination. Extensive experiments are conducted on six different datasets to predict the information dissemination range and the dissemination trend of the social network. The experimental results demonstrate that the proposed algorithm outperforms traditional information dissemination models in simulating the information dissemination process in current social networks, providing valuable insights for early warning of adverse social events.

\section{Acknowledge}
This work is supported by the National Natural Science Foundation of China (Grant No. 62302145), the key program of Anhui Province Key Laboratory of Affective Computing and Advanced Intelligent Machine (Grant No. PA2023GDSK0059), and Young teachers’ scientific research innovation launches special A project (Grant No. JZ2023HGQA0100).



% \bibliographystyle{unsrt}
% \bibliography{con}

\bibliographystyle{IEEEtran}
\bibliography{con}
\end{document}